\documentstyle{aipproc}

\newcommand{\ud}{\mathrm{d}}
\begin{document}
\title{Symmetry, Hamiltonian Problems and Wavelets in Accelerator Physics}
\author{A.~Fedorova$^*$,  M.~Zeitlin$^*$ and Z.~Parsa$^\dagger$}
\address{$^*$Institute of Problems of Mechanical Engineering,
 Russian Academy of Sciences, 199178, Russia, St.~Petersburg,
  V.O., Bolshoj pr., 61, e-mail: zeitlin@math.ipme.ru and\\
 $^\dagger$Dept. of Physics, Bldg.~901A, Brookhaven National Laboratory,
Upton, NY 11973-5000, USA}
\maketitle

\begin{abstract}
In this paper we consider applications of methods from wavelet analysis to
nonlinear dynamical problems related to accelerator physics. In our approach we
take into account underlying algebraical, geometrical and topological
structures of corresponding problems.
\end{abstract}
\section{Introduction}
This paper is the sequel of our first paper in this volume [1],
in which we considered the applications of a number of analytical methods from
nonlinear (local) Fourier analysis, or wavelet analysis, to nonlinear
accelerator physics problems. This paper is the continuation of results
from [2]--[7], which is based on our approach to investigation of nonlinear
problems both general and with additional structures (Hamiltonian, symplectic
or quasicomplex), chaotic, quasiclassical, quantum.

 Wavelet analysis is
a relatively novel set of mathematical methods, which gives us a possibility
to work with well-localized bases in functional spaces and with the
general type of operators (differential, integral, pseudodifferential) in
such bases.

 In contrast  with paper [1] in this paper we try to take into account
before using power analytical approaches underlying algebraical, geometrical,
topological structures related to kinematical, dynamical and hidden
symmetry of physical problems. In this paper we give a review of
a number of the corresponding problems and describe the key points of
some possible methods by which we can find the full solutions of initial
physical problem. We described  a few concrete problems in [1, part II].
The most interesting case is the dynamics of spin-orbital motion [1, II D].
Related problems may be found in [8].

The content of this paper is not more than an attempt to extract the most
complicated formal or mathematical or principal parts of the World of
nonlinear accelerator physics, which is today beyond of mainstream in our opinion.

In part II we consider dynamical consequences of covariance properties
regarding to relativity (kinematical) groups and continuous wavelet transform
as a method for the solution of dynamical problems.

In part II A we introduce the semidirect product structure, which allows us to
consider from general point of view all relativity groups such as Euclidean,
Galilei, Poincare.

Then in part II B we consider the Lie-Poisson equations and
obtain the manifestation of semiproduct structure of (kinematic) symmetry group
on dynamical level. So, correct description of dynamics is a consequence of
correct understanding of real symmetry of the concrete problem.

In part II C we consider the technique for simplifications of dynamics related
to semiproduct structure by using reduction to corresponding orbit structure.
As result we have simplified Lie-Poisson equations.

In part II D we consider the Lagrangian theory related to semiproduct structure
and explicit form of variation principle and corresponding (semidirect)
Euler-Poicare equations.

In part II E we introduce a continuous wavelet transform and corresponding
analytical technique which allow to consider covariant wavelet analysis.

In part II F we consider in the particular case of affine Galilei group
with the semiproduct structure also the
corresponding orbit technique for constructing different types of invariant
wavelet bases.

 In part III we consider instead of kinematical symmetry the
dynamical symmetry.

In part III A
according to the orbit method and by using construction
from the geometric quantization theory we construct the symplectic
and Poisson structures associated with generalized wavelets by
using metaplectic structure. We consider wavelet approach to
the calculations of Melnikov functions in the theory of homoclinic
chaos in perturbed Hamiltonian systems in part III B and for calculation of
Arnold--Weinstein curves (closed loops) in Floer variational approach in part III C.

In parts III D, III E  we consider applications of very
useful fast wavelet transform technique (part III F) to calculations in
symplectic scale of spaces and to quasiclassical evolution dynamics.
This method gives maximally sparse representation of (differential) operator
 that allows us to take into account contribution from each level of
resolution.

 In part IV A we consider symplectic and Lagrangian structures for the case of
discretization of flows by corresponding maps and  in part IV B construction of
corresponding solutions by applications of
generalized wavelet approach which is based on generalization of
  multiresolution analysis
for the case of maps.
\section{Semidirect Product, Dynamics, Wavelet Representation}
\subsection{Semidirect Product}
Relativity groups such as Euclidean, Galilei or Poincare groups  are the
particular cases of semidirect product construction, which is very useful and
simple general construction in the group theory [9]. We may consider as a basic
example the Euclidean group $SE(3)=SO(3)\bowtie{\bf R}^3$, the semidirect
product of rotations and translations. In general case we have $S=G\bowtie V$,
where group G (Lie group or automorphisms group) acts on a vector space V and
on its dual $V^*$.
Let $V$ be a vector space and $G$ is the Lie group, which acts on the left by
linear maps on V (G also acts on the left on its dual space $V^*$).
The semidirect product $S=G\bowtie V$ is the Cartesian product
$S=G\times V$ with group multiplication
\begin{equation}
(g_1,v_1)(g_2,v_2)=(g_1g_2,v_1+g_1v_2),
\end{equation}
where the action of $g\in G$ on $v\in V$ is denoted as $gv$. Of course, we
can consider the corresponding definitions both in case of
the right actions and in case, when G is a group of automorphisms of the vector
space V. As we shall explain below both cases, Lie groups and automorphisms
groups, are important for us.

So, the Lie algebra of S is the semidirect product  Lie algebra,
$s={\mathcal{G}} \bowtie V$ with brackets
\begin{equation}
[(\xi_1,v_1),(\xi_2,v_2)]=([\xi_1,\xi_2],\xi_1v_2-\xi_2v_1),
\end{equation}
where the induced action of $\mathcal{G}$ by concatenation is denoted
as $\xi_1 v_2$. Also we need expressions for adjoint and coadjoint actions for
semidirect products. Let $(g,v)\in S=G\times V, \quad (\xi,u)\in
s={\mathcal{G}}\times V$, $(\mu,a)\in s^*={\mathcal G}^*\times V^*$, $g\xi=Ad_g\xi$,
$g\mu=Ad^*_{g^{-1}}\mu$, $ga$ denotes the induced left action of $g$ on $a$ (the
left action of G on V induces a left action on $V^*$ --- the inverse of the
transpose of the action on V), $\rho_v: {\cal G}\to V$ is a linear map
given by $\rho_v(\xi)=\xi v$, $\rho^*_v: V^*\to{\cal G}^*$ is its dual.
Then these actions are given by simple concatenation:
\begin{eqnarray}
(g,v)(\xi,u)&=&(g\xi,gu-(g\xi)v),\\
(g,v)(\mu,a)&=&(g\mu+\rho^*_v(ga),ga)\nonumber
\end{eqnarray}
Below we use the following notation: $\rho^*_v a=v\diamond a\in{\cal G^*}$ for
$a\in V^*$, which is a bilinear operation in $v$ and $a$. So, we have the coadjoint
action:
\begin{equation}
(g,v)(\mu,a)=(g\mu+v\diamond(ga),ga).
\end{equation}
Using concatenation notation for Lie algebra actions we have alternative
definition of $v\diamond a\in{\mathcal G}^*$.
For all $v\in V$, $a\in V^*$, $\eta\in{\mathcal G}$ we have
\begin{equation}
<\eta a,v>=-<v\diamond a, \eta>
\end{equation}

\subsection{The Lie-Poisson Equations and Semiproduct Structure}
Below we consider the manifestation of semiproduct structure
of symmetry group on dynamical level.
Let $F,G$ be real valued functions on the dual space ${\mathcal G}^*$,
 $\mu\in{\mathcal G}^*$. Functional
derivative of F at $\mu$ is the unique element $\delta F/\delta\mu\in{\mathcal
G}$:
\begin{equation}\label{eq:lim}
\lim_{\epsilon\to 0}\frac{1}{\epsilon}
\lbrack F(\mu+\epsilon\delta\mu)-F(\mu)\rbrack=
<\delta\mu,\frac{\delta F}{\delta\mu}>
\end{equation}
for all $\delta\mu\in{\mathcal G}^*$, $<,>$ is pairing between $\mathcal G^*$ and
$\mathcal G$.

Define the $(\pm)$ Lie-Poisson brackets by
\begin{equation}\label{eq:FG}
\{F,G\}_\pm(\mu)=\pm <\mu,\lbrack\frac{\delta F}{\delta\mu},
\frac{\delta G}{\delta\mu}\rbrack>
\end{equation}
The Lie-Poisson equations, determined by
\begin{equation}
\dot{F}=\{F,H\}
\end{equation}
read intrinsically
\begin{equation} \label{eq:mu}
\dot{\mu}=\mp ad^*_{\partial H/\partial\mu}\mu.
\end{equation}
For the left representation of G on V $\pm$ Lie-Poisson bracket of two
functions $f,k: s^*\to {\bf R}$ is given by
\begin{equation}\label{eq:fk}
\{f,k\}_{\pm}(\mu, a)=\pm <\mu,\lbrack\frac{\delta f}{\delta\mu},
\frac{\delta k}{\delta\mu}\rbrack>\pm
<a,\frac{\delta f}{\delta\mu}\frac{\delta k}{\delta a}-
\frac{\delta k}{\delta\mu}\frac{\delta f}{\delta a}>,
\end{equation}
where $\delta f/\delta\mu\in{\mathcal G}$, $\delta f/\delta a\in V$ are
the functional derivatives of f (\ref{eq:lim}). The Hamiltonian vector field of
$h: s^*\in{\bf R}$ has the expression
\begin{equation}
X_h(\mu,a)=\mp(ad^*_{\delta h/\delta\mu}\mu-\frac{\delta h}{\delta
a}\diamond a, -\frac{\delta h}{\delta\mu}a).
\end{equation}
Thus, Hamiltonian equations on the dual of a semidirect product are [9]:
\begin{eqnarray}\label{eq:mua}
\dot{\mu}&=&\mp ad^*_{\delta h / \delta\mu}\mu\pm\frac{\delta h}{\delta a}\diamond
 a\\
\dot{a}&=&\pm\frac{\delta h}{\delta\mu} a \nonumber
\end{eqnarray}
So, we can see the explicit difference between Poisson brackets (\ref{eq:FG})
and (\ref{eq:fk}) and the equations of motion (\ref{eq:mu}) and (\ref{eq:mua}),
which come from the semiproduct structure.

\subsection{Reduction of Dynamics on Semiproduct}
There is technique for reducing dynamics that is associated with the geometry
of semidirect product reduction theorem[9].
Let us have a Hamiltonian on $T^*G$ that is invariant under the isotropy
$G_{a_0}$ for $a_0\in V^*$. The semidirect product reduction theorem states
that reduction  of $T^* G$ by $G_{a_0}$ gives reduced spaces that are
simplectically diffeomorphic to coadjoint orbits in the dual of the Lie algebra
of the semidirect product $({\mathcal G}\bowtie V)^*$.
If one reduces the semidirect group product $S=G\bowtie V$ in two stages, first
by V and then by G one recovers this semidirect product reduction theorem.
Thus, let $S=G\bowtie V$, choose $\sigma=(\mu,a)\in{\mathcal G}^*\times V^*$
and reduce $T^*S$ by the action of $S$ at $\sigma$ giving the coadjoint orbit
${\mathcal O}_\sigma$ through $\sigma\in S^*$. There is a symplectic
diffeomorphism between ${\mathcal O}_\sigma$ and the reduced space obtained be
reducing $T^*G$ by the subgroup $G_a$ (the isotropy of G for its action on
$V^*$ at the point $a\in V^*$) at the point $\mu|{\mathcal G}_a$, where
${\mathcal G}_a$ is the Lie algebra of $G_a$.

 Then
we have the following procedure.
\begin{enumerate}
\item We start with a Hamiltonian $H_{a_0}$ on $T^*G$ that depends
parametrically on a variable $a_0\in V^*$.
\item The Hamiltonian regarded as a map: $T^*G\times V^*\to{\bf R}$ is
assumed to be invariant on $T^*G$ under the action of G on $T^*G\times V^*$.
\item The condition 2 is equivalent to the invariance of the function H defined
on $T^*S=T^*G\times V\times V^*$ extended to be constant in the variable  V
under the action of the semidirect product.
\item By the semidirect product reduction theorem,
 the dynamics of $H_{a_0}$ reduced by
$G_{a_0}$, the isotropy group of $a_0$, is simplectically equivalent to
Lie-Poisson dynamics on $s^*={\mathcal G}^*\times V^*$.
\item This Lie-Poisson dynamics is given by equations (\ref{eq:mua}) for the
function $h(\mu,a)=H(\alpha_g,g^{-1}a)$, where $\mu=g^{-1}\alpha_g$.
\end{enumerate}

\subsection{Lagrangian Theory, the Euler-Poincare Equations, Variational
Approach on Semiproduct}
Now we consider according to [9] Lagrangian side of a theory.
This approach is based on variational principles with symmetry and is not
dependent on Hamiltonian formulation, although it is demonstrated in [9] that this
purely Lagrangian formulation is equivalent to the Hamiltonian formulation
on duals of semidirect product (the
corresponding Legendre transformation is a diffeomorphism).

We consider the case of the left representation and
the left invariant  Lagrangians ($\ell$ and L),
which depend in additional on another parameter $a\in V^*$ (dynamical
parameter),
where V is  representation space for the Lie group G and L has  an invariance
property related to both arguments. It should be noted that the resulting
equations of motion, the Euler-Poincare equations, are not the Euler-Poincare
equations for the semidirect product Lie algebra ${\mathcal G}\bowtie V^*$ or
${\mathcal G}\bowtie V$.

So, we have the following:
\begin{enumerate}
\item There is a left presentation of Lie group G on the vector space V and G
acts in the natural way on the left on $TG\times V^*: h(v_g,a)=(hv_g,ha)$.
\item The function $L: TG\times V^*\in{\bf R}$ is the left G-invariant.
\item Let $a_0\in V^*$, Lagrangian $L_{a_0}: TG\to{\bf R}$,
$L_{a_0}(v_g)=L(v_g,a_0)$. $L_{a_0}$ is left invariant under the lift to TG of
the left action of $G_{a_0}$ on G, where $G_{a_0}$ is the isotropy group of
$a_0$.
\item Left G-invariance of L permits us to define
\begin{equation}
\ell:{\mathcal G}\times V^*\to{\bf R}
\end{equation}
by
\begin{equation}
\ell(g^{-1}v_g,g^{-1}a_0)=L(v_g,a_0).
\end{equation}
This relation defines for any $\ell:{\mathcal G}\times V^*\to{\bf R}$ the left
G-invariant function $L: TG\times V^*\to{\bf R}$.
\item For a curve $g(t)\in G$ let be
\begin{equation}
\xi(t):=g(t)^{-1}\dot{g}(t)
\end{equation}
and define the curve $a(t)$ as the unique solution of the following linear
differential equation with time dependent coefficients
\begin{equation}
\dot{a}(t)=-\xi(t)a(t),
\end{equation}
with initial condition $a(0)=a_0$. The solution can be written as
$a(t)=g(t)^{-1}a_0$.
\end{enumerate}
Then we have four equivalent descriptions of the corresponding dynamics:
\begin{enumerate}
\item If $a_0$ is fixed then Hamilton's variational principle
\begin{equation}
\delta\int_{t_1}^{t_2}L_{a_0}(g(t),\dot{g}(t){\rm d}t=0
\end{equation}
holds for variations $\delta g(t)$ of $g(t)$ vanishing at the endpoints.
\item $g(t)$ satisfies the Euler-Lagrange equations for $L_{a_0}$ on G.
\item The constrained variational principle
\begin{equation}
\delta\int_{t_1}^{t_2}\ell(\xi(t), a(t)){\rm d}t=0
\end{equation}
holds on ${\mathcal G}\times V^*$, using variations of $\xi$ and $a$ of the form
 $\delta\xi=\dot{\eta}+[\xi,\eta]$, $\delta a=-\eta a$, where
$\eta(t)\in{\mathcal G}$ vanishes at the endpoints.
\item The Euler-Poincare equations hold on ${\mathcal G}\times V^*$
\begin{equation}
\frac{{\rm d}}{{\rm d}t}\frac{\delta\ell}{\delta\xi}=
ad_\xi^*\frac{\delta\ell}{\delta\xi}+\frac{\delta\ell}{\delta a}\diamond a
\end{equation}
\end{enumerate}
So, we may apply our wavelet methods either on the level of variational formulation
(17) or on the level of Euler-Poincare equations (19).

\subsection{Continuous Wavelet Transform}
  Now we need take into account the Hamiltonian
or Lagrangian structures related with systems (12) or  (19).
Therefore, we need to consider generalized wavelets, which
allow us to consider the corresponding  structures  instead of
compactly supported wavelet representation from paper [1].

In wavelet analysis the following three concepts are used now:
1).\ a square integrable representation $U$ of a group $G$,
2).\ coherent states (CS) over G,
3).\ the wavelet transform associated to U. We consider now
their unification [10], [11].

Let $G$ be a locally compact group and $U_a$  strongly continuous,
irreducible, unitary representation of G on Hilbert space ${\mathcal H}$.
Let $H$ be a closed subgroup of $G$, $X=G/H$ with (quasi) invariant measure $\nu$
and $\sigma: X=G/H\to G$ is a Borel section in a principal bundle $G\to G/H$.
Then we say that $U$ is square integrable $mod(H,\sigma)$ if there exists a
non-zero vector $\eta\in{\mathcal H}$ such that
\begin{equation}\label{eq:int}
0<\int_X|<U(\sigma(x))\eta|\Phi>|^2{\rm d}\nu(x)=<\Phi|A_\sigma\Phi>\ <\infty,
\quad \forall\Phi\in{\mathcal H}
\end{equation}
Given such a vector $\eta\in{\mathcal H}$ called admissible for $(U,\sigma)$ we
define the family of (covariant) coherent states or wavelets, indexed by points
$x\in X$, as the orbit of $\eta$ under $G$, though the representation $U$ and the
section $\sigma$ [10], [11]
\begin{equation}
S_\sigma={\eta_{\sigma(x)}=U(\sigma(x))\eta|x\in X}
\end{equation}
So, coherent states or wavelets are simply the elements of the orbit under U of
a fixed vector $\eta$ in representation space.
We have the following fundamental properties:
\begin{enumerate}
\item Overcompleteness: \\
 The set $S_\sigma$ is total in ${\mathcal H}:(S_\sigma)^\perp={0}$
\item Resolution property:\\
the square integrability condition (\ref{eq:int}) may be represented as a
resolution relation:
\begin{equation}
\int_X|\eta_\sigma(x)><\eta_{\sigma(x)}|{\rm d}\nu(x)=A_\sigma,
\end{equation}
where $A_\sigma$ is a bounded, positive operator with a densely defined
inverse. Define the linear map
\begin{equation}
W_\eta: {\mathcal H}\to L^2(X,{\rm
d}\nu),(W_\eta\Phi)(x)=<\eta_{\sigma(x)}|\Phi>
\end{equation}
Then the range $H_\eta$ of $W_\eta$ is complete with respect to the scalar
product $<\Phi|\Psi>_\eta=<\Phi|W_\eta A^{-1}_\sigma W^{-1}_\eta\Psi>$ and $W_\eta$ is
unitary operator from ${\mathcal H}$ onto ${\mathcal H}_\eta$.
$W_\eta$ is Continuous Wavelet Transform (CWT).
\item Reproducing kernel\\
The orthogonal projection from $L^2(X,{\rm d}\nu)$ onto ${\mathcal H}_\eta$ is
an integral operator $K_\sigma$ and $H_\eta$ is a reproducing kernel Hilbert
space of functions:
\begin{equation}
\Phi(x)=\int_XK_\sigma(x,y)\Phi(y){\rm d}\nu(y), \quad \forall \Phi\in{\mathcal
H}_\eta.
\end{equation}
The kernel is given explicitly by
$K_\sigma(x,y)=<\eta_{\sigma(x)}|A_\sigma^{-1}\eta_{\sigma(y)}>$, if
$\eta_{\sigma(y)}\in D(A^{-1}_\sigma)$, $\forall y\in X$.
So, the function $\Phi\in L^2(X,{\rm d}\nu)$ is a wavelet transform (WT)
iff it satisfies this reproducing relation.
\item Reconstruction formula.\\
The WT $W_\eta$ may be inverted on its range by the adjoint operator,
$W_\eta^{-1}=W_\eta^*$ on ${\mathcal H}_\eta$ to obtain for
$\eta_{\sigma(x)}\in D(A_\sigma^{-1})$, $\forall x\in X$
\begin{equation}
W_\eta^{-1}\Phi=\int_X\Phi(x)A_\sigma^{-1}\eta_{\sigma(x)}{\rm d}\nu(x), \quad
\Phi\in{\mathcal H}_\eta.
\end{equation}
This is inverse WT.
\end{enumerate}
If $A_\sigma^{-1}$ is bounded then $S_\sigma$ is called a frame, if
$A_\sigma=\lambda I$ then $S_\sigma$ is called a tight frame. This two cases
are generalization of a simple case, when $S_\sigma$ is an (ortho)basis.

The most simple cases of this construction are:\\
1. $H=\{e\}$. This is the standard construction of WT over a locally compact
group. It should be noted that the square integrability of U is equivalent to
U belonging to the discrete series. The most simple example is related to
 the affine $(ax+b)$ group and yields the usual one-dimensional wavelet
analysis
\begin{equation}
[\pi(b,a)f](x)=\frac{1}{\sqrt{a}}f\left(\frac{x-b}{a}\right).
\end{equation}
For $G=SIM(2)={\bf R}^2\bowtie({\bf R}^{+}_*\times SO(2))$,
the similitude group of the plane, we have the corresponding two-dimensional
wavelets.

2. $H=H_\eta$, the isotropy (up to a phase) subgroup of $\eta$: this is the
case of the Gilmore-Perelomov CS. Some cases of group G are:\\
a). Semisimple groups, such as SU(N), SU(N$|$M), SU(p,q), Sp(N,{\bf R}).\\
b). the Weyl-Heisenberg  group $G_{WH}$ which leads to the Gabor
functions, i.e. canonical (oscillator)coherent states associated
with windowed Fourier
transform or Gabor transform (see also part III A):
\begin{equation}
[\pi(q,p,\varphi)f](x)=\exp(i\mu(\varphi-p(x-q))f(x-q)
\end{equation}
In this case H is the  center of $G_{WH}$.
In both cases time-frequency plane corresponds to the phase
space of group representation.\\
c). The similitude group SIM(n) of ${\bf R}^n(n\ge3)$: for $H=SO(n-1)$ we have
the axisymmetric n-dimensional wavelets.\\
d). Also we have
 the case of bigger group, containing
both affine and We\-yl-\-Hei\-sen\-berg group, which interpolate between
affine wavelet analysis and windowed Fourier analysis: affine
Weyl--Heisenberg group [11].\\
e). Relativity groups. In a nonrelativistic setup, the natural kinematical
group is the (extended) Galilei group. Also we may adds independent space and
time dilations and obtain affine Galilei group. If we restrict the dilations by
the relation $a_0=a^2$, where $a_0, a$ are the time and space dilation we
obtain the Galilei-Schr\"odinger group, invariance group of both Schr\"odinger
and heat equations. We consider these examples in the next section. In the same
way we may consider as kinematical group the Poincare group. When $a_0=a$ we
have affine Poincare or Weyl-Poincare group. Some useful generalization of that
affinization construction we consider for the case of hidden metaplectic structure
 in section III A.

But the usual representation  is not
square--integrable and must be modified: restriction of the
representation to a suitable quotient space of the group (the
associated phase space in our case) restores square --
integrability: $G\longrightarrow$ homogeneous space.\\
Also, we have more general approach which allows to consider wavelets
corresponding to more general groups and representations [12], [13].

Our goal is applications of these results to problems of
Hamiltonian dynamics and as consequence we need to take into account
symplectic nature of our dynamical problem.
 Also, the symplectic and wavelet structures
 must be consistent (this must
be resemble the symplectic or Lie-Poisson integrator theory).
 We use the
point of view of geometric quantization theory (orbit method)
instead of harmonic analysis. Because of this we can consider
(a) -- (e) analogously.

\subsection{Bases for Solutions}
We consider an important particular case of affine
relativity group (relativity group
combined with dilations) --- affine Galilei group in n-dimensions. So, we have
combination of Galilei group with independent space and time dilations:
$G_{aff}=G_m\bowtie D_2$,
where $D_2=({\bf R}^{+}_*)^2\simeq {\bf R}^2$, $G_m$ is extended
Galilei group corresponding to mass parameter $m>0$ ($G_{aff}$ is noncentral
extension of $G\bowtie D_2$ by ${\bf R}$, where G is usual Galilei group).
Generic element of $G_{aff}$ is $g=(\Phi,b_0,b;v;R,a_0,a)$, where
$\Phi\in{\bf R}$ is the extension parameter in $G_m$, $b_0\in{\bf R}$,
$b\in{\bf R}^n$ are the time and space translations, $v\in{\bf R}^n$ is the boost
parameter, $R\in SO(n)$ is a rotation and $a_0,a\in{\bf R}^+_*$ are time and
space dilations. The actions of $g$ on space-time  is then $x\mapsto
aRx+a_0vt+b$, $t\mapsto a_0t+b_0$, where $x=(x_1,x_2,...,x_n)$.
The group law is
\begin{eqnarray}
gg'=&&(\Phi+\frac{a^2}{a_0}\Phi'+avRb'+\frac{1}{2}a_0v^2b'_0, b_0+a_0b'_0,
       b+aRb'+a_0vb'_0;\\
    && v+\frac{a}{a_0}Rv', RR'; a_aa'_0, aa')\nonumber
\end{eqnarray}
It should be noted that $D_2$ acts nontrivially on $G_m$.
Space-time wavelets associated to $G_{aff}$ corresponds to unitary irreducible
representation of spin zero. It may be obtained via orbit method. The Hilbert
space is ${\mathcal H}=L^2 ({\bf R}^n\times{\bf R},
{\rm d}k{\rm d}\omega)$, $k=(k_1,...,k_n)$, where ${\bf R}^n\times{\bf R}$ may
be identified with usual Minkowski space and we have for representation:
\begin{equation}
(U(g)\Psi)(k,\omega)=\sqrt{a_0a^n}{\rm exp}i(m\Phi+kb-\omega
b_0)\Psi(k',\omega'),
\end{equation}
with  $k'=aR^{-1}(k+mv)$, $\omega'=a_0(\omega-kv-\frac{1}{2}mv^2)$,
$m'=(a^2/a_0) m$.
Mass m is a coordinate in the dual of the Lie algebra and these relations are a
part of coadjoint action  of $G_{aff}$. This representation is unitary and
irreducible but not square integrable. So, we need to consider reduction to the
corresponding quotients $X=G/H$. We consider the case in which H=\{phase
changes $\Phi$ and space dilations $a$\}. Then the space $X=G/H$ is parametrized
by points $\bar{x}=(b_0,b;v;R;a_0)$. There is a dense set of vectors
$\eta\in{\mathcal H}$ admissible ${\rm mod}(H,\sigma_\beta)$, where
$\sigma_\beta$ is the
corresponding section.
We have a two-parameter family of functions $\beta$(dilations):
$\beta(\bar{x})=(\mu_0+\lambda_)a_0)^{1/2}$, $\lambda_0, \mu_0\in{\bf R}$.
Then any admissible vector $\eta$ generates a tight frame of Galilean wavelets
\begin{equation}
\eta_{\beta(\bar{x})}(k,\omega)=\sqrt{a_0(\mu_0+\lambda_0a_0)^{n/2}}
{\rm e}^{i(kb-\omega b_0)}\eta(k',\omega'),
\end{equation}
with $k'=(\mu_0+\lambda_0 a)^{1/2}R^{-1}(k+mv)$,
$\omega'=a_0(\omega-kv-mv^2/2)$.
The simplest examples of admissible vectors (corresponding to usual Galilei
case) are Gaussian vector: $\eta(k)\sim{\rm exp}(-k^2/2mu)$ and
binomial vector: $\eta(k)\sim(1+k^2/2mu)^{-\alpha/2}$, $\alpha> 1/2$, where $u$
is a kind of internal energy.
When we impose the relation $a_0=a^2$ then we have the restriction to the
Galilei-Schr\"odinger group $G_s=G_m\bowtie D_s$, where $D_s$ is the
one-dimensional subgroup of $D_2$. $G_s$ is a natural invariance group of
both the Schr\"odinger equation and the heat equation.
The restriction to $G_s$ of the representation (29) splits into the direct
sum of two irreducible ones $U=U_+\oplus U_-$ corresponding to the
decomposition $L^2({\bf R}^n\times{\bf R}, {\rm d}k{\rm d}\omega)=
{\mathcal H}_+\oplus{\mathcal H}_-$, where
\begin{eqnarray}
{\mathcal H}_\pm&=&L^2(D_{\pm}, {\rm d}k{\rm d}\omega\\
&=& \{ \psi\in L^2({\bf R}^n\times{\bf R},{\rm d}k{\rm d}\omega),\quad
\psi(k,\omega)=0\quad {\textrm for} \quad \omega+k^2/2m=0\}\nonumber
\end{eqnarray}
These two subspaces are the analogues of usual Hardy spaces on
${\bf R}$, i.e. the subspaces of (anti)progressive wavelets (see also below,
part III A).
The two representation $U_\pm$ are square integrable modulo the center.
There is a dense set of admissible vectors $\eta$, and each of them generates a
set of $CS$ of Gilmore-Perelomov type.
Typical wavelets of this kind are:\\
the Schr\"odinger-Marr wavelet:
\begin{equation}
\eta(x,t)=(i\partial_t+\frac{\triangle}{2m}){\rm e}^{-(x^2+t^2)/2}
\end{equation}
the Schr\"odinger-Cauchy wavelet:
\begin{equation}
\psi(x,t)=(i\partial_t+\frac{\triangle}{2m})\frac{1}
{(t+i)\prod_{j=1}^n(x_j+i)}
\end{equation}
So, in the same way we can construct  invariant bases with explicit
manifestation of underlying symmetry for solving Hamiltonian (12) or Lagrangian
(19) equations.

\section{Symplectic Structures, Quantization and Fast Wavelet Transform}
\subsection{Metaplectic Group and Representations}
Let $Sp(n)$ be
symplectic group, $Mp(n)$ be its unique two- fold covering --
metaplectic group [14].   Let V be a symplectic vector space
with symplectic form ( , ), then $R\oplus V$ is nilpotent Lie
algebra - Heisenberg algebra:
$$[R,V]=0, \quad [v,w]=(v,w)\in
R,\quad [V,V]=R.$$
$Sp(V)$ is a group of automorphisms of
Heisenberg algebra.

 Let N be a group with Lie algebra $R\oplus
V$, i.e.  Heisenberg group.  By Stone-- von Neumann theorem
Heisenberg group has unique irreducible unitary representation
in which $1\mapsto i$. Let us also consider the  projective
representation of simplectic group $Sp(V)$:
$U_{g_1}U_{g_2}=c(g_1,g_2)\cdot U_{g_1g_2}$, where c is a map:
$Sp(V)\times Sp(V)\rightarrow S^1$, i.e. c  is $S^1$-cocycle.

But this representation is unitary representation of universal
covering, i.e. metaplectic group $Mp(V)$. We give this
representation without Stone-von Neumann theorem.\
Consider a new group $F=N'\bowtie Mp(V),\quad \bowtie$ is semidirect
product (we consider instead of  $ N=R\oplus V$ the $
N'=S^1\times V, \quad S^1=(R/2\pi Z)$). Let $V^*$ be dual to V,
$G(V^*)$ be automorphism group of $V^*$.Then F is subgroup of $
G(V^*)$, which consists of elements, which acts on $V^*$ by affine
transformations. \\
This is the key point!

 Let $q_1,...,q_n;p_1,...,p_n$ be symplectic basis in V,
$\alpha=pdq=\sum p_{i}dq_i $  and $d\alpha$ be symplectic form on
$V^*$. Let M be fixed affine polarization, then for $a\in F$ the
map $a\mapsto \Theta_a$ gives unitary representation of G:
$ \Theta_a: H(M) \rightarrow H(M) $

Explicitly  we have for representation of N on H(M):
$$
(\Theta_qf)^*(x)=e^{-iqx}f(x),  \quad
 \Theta_{p}f(x)=f(x-p)
$$
The representation of N on H(M) is irreducible. Let $A_q,A_p$
be infinitesimal operators of this representation
$$
 A_q=\lim_{t\rightarrow 0} \frac{1}{t}[\Theta_{-tq}-I], \quad
 A_p=\lim_{t\rightarrow 0} \frac{1}{t}[\Theta_{-tp}-I],
$$
$$\mbox{then}\qquad
A_q f(x)=i(qx)f(x),\quad A_p f(x)=\sum p_j\frac{\partial
f}{\partial x_j}(x)
$$
Now we give the representation of infinitesimal ba\-sic
elements. Lie algebra of the group F is the algebra of all
(non\-ho\-mo\-ge\-ne\-ous) quadratic po\-ly\-no\-mi\-als of (p,q) relatively
Poisson bracket (PB). The basis of this algebra consists of
elements
$1,q_1,...,q_n$,\  $p_1,...,p_n$,\ $ q_i q_j, q_i p_j$,\ $p_i p_j,
 \quad i,j=1,...,n,\quad i\leq j$,
 \begin{eqnarray*}
 & &PB \ is
\quad \{ f,g\}=\sum\frac{\partial f}{\partial p_j}
\frac{\partial g}{\partial q_i}-\frac{\partial f}{\partial q_i}
\frac{\partial g}{\partial p_i} \quad
 \mbox{and}  \quad
   \{1,g \}= 0 \quad for \mbox{ all} \ g,\\
& &\{ p_i,q_j\}= \delta_{ij},\quad \{p_i
q_j,q_k\}=\delta_{ik}q_j,\quad
    \{p_i q_j,p_k\}=-\delta_{jk}p_i, \quad \{p_ip_j,p_k\}=0,\\
& & \{p_i p_j,q_k \}=
\delta_{ik}p_j+\delta_{jk}p_i,\quad
  \{ q_i q_j,q_k\}=0,
 \{q_i q_j,p_k\}=-\delta_{ik}q_j-\delta_{jk}q_i
 \end{eqnarray*}
so, we have the representation of basic elements
 $ f\mapsto A_f : 1\mapsto i, q_k\mapsto ix_k $,
\begin{eqnarray*}
 p_l\mapsto\frac{\delta}{\delta x^l}, p_i q_j\mapsto
x^i\frac{\partial}{\partial x^j}+\frac{1}{2}\delta_{ij},\qquad
 p_k p_l\mapsto \frac{1}{i}\frac{\partial^k}{\partial x^k\partial
x^l}, q_k q_l\mapsto ix^k x^l
\end{eqnarray*}
This gives  the structure of the Poisson mani\-folds to
representation of any (nilpotent) algebra or in other words to
continuous wavelet trans\-form.

{\bf  The Segal-Bargman Representation.}
Let $ z=1/\sqrt{2}\cdot(p-iq),\quad
\bar{z}=1/\sqrt{2}\cdot(p+iq),\quad
$
$ p=(p_1,...,p_n)
,\quad
 F_n $ is the
space of holomorphic functions of n complex variables with
$(f,f)< \infty$, where $$ (f,g)=(2\pi)^{-n}\int
f(z)\overline{g(z)}e^{-|z|^2}dpdq $$
Consider a map  $U:
H\rightarrow F_n$ , where H is with real polarization, $F_n
$ is with complex polarization, then we have $$(U\Psi)(a)=\int
A(a,q)\Psi(q)dq,\qquad \mbox{where}\quad
A(a,q)=\displaystyle\pi^{-n/4}e^{-1/2(a^2+q^2)+\sqrt{2}aq}
$$
i.e. the Bargmann formula produce  wavelets.We also have
the representation of Heisenberg algebra on $F_n$ :
\begin{eqnarray*}
U\frac{\partial}{\partial q_j} U^{-1}=\frac{1}{\sqrt{2}}\left
(z_j- \frac{\partial}{\partial z_j}\right),\qquad
 Uq_j
U^{-1}=-\frac{i}{\sqrt{2}}\left(z_j+\frac{\partial }{\partial
 z_{j}} \right)
\end{eqnarray*}
and also : $ \omega=d\beta=dp\wedge dq,$
 where
$\beta=i\bar{z}dz $.

{\bf Orbital Theory for  Wavelets.}
Let coadjoint action be
$<g\cdot f,Y>=<f,Ad(g)^{-1}Y>,$
 where $<,>$ is pairing
$ g\in G,\quad f\in g^*,\quad Y\in{\cal G}$.
The orbit is
${\cal O}_f=G\cdot f\equiv G/G(f)$.
Also, let A=A(M) be algebra of functions,
V(M) is A-module of vector fields,
$A^p$  is A-module of p-forms.
Vector fields on orbit is
$$
\sigma({\cal O},X)_f(\phi)=\frac{d}{dt}(\phi(\exp tXf))\Big |_{t=0}
$$
where $\phi\in A({\cal O}),\quad f\in{\cal O}$. Then ${\cal O}_f$
are homogeneous symplectic manifolds with 2-form
$
\Omega(\sigma({\cal O},X)_f,\sigma({\cal  O},Y)_f)=<f,[X,Y]>,
$
and $d\Omega=0$. PB on ${\cal O}$ have the next form
$
\{ \Psi_1,\Psi_2\}=p(\Psi_1)\Psi_2
$
where p is $ A^1({\cal O})\rightarrow V({\cal O})$ with
definition
$\Omega (p(\alpha),X)$ $=$ $i(X)\alpha$. Here $\Psi_1,\Psi_2\in
A(\cal {O})$ and  $A({\cal O}) $ is Lie algebra with bracket
\{,\}.
Now let N be a Heisenberg group. Consider adjoint and
coadjoint representations in some particular case.
 $N=(z,t)\in C\times R,
 z=p+iq$; compositions in N are $(z,t)\cdot(z',t')=
 (z+z',t+t'+B(z,z')) $, where $B(z,z')=pq'-qp'$. Inverse
 element is $(-t,-z)$. Lie algebra n of N is  $(\zeta,\tau)
 \in C\times R$ with bracket $[(\zeta,\tau),(\zeta',\tau')]=
 (0,B(\zeta,\zeta'))$. Centre is $\tilde{z}\in n $ and
generated by (0,1);
 Z is a subgroup $\exp\tilde{z}$.
Adjoint representation N on n is given by formula
  $
Ad(z,t)(\zeta,\tau)=(\zeta,\tau+B(z,\zeta))
$
Coadjoint:
 for $f\in n^*,\quad g=(z,t)$,
$(g \cdot f)(\zeta,\zeta)=f(\zeta,\tau)-B(z,\zeta)f(0,1)$ then
  orbits for which $f|_{\tilde z}\neq 0$ are plane in $n^*$
 given by equation $ f(0,1)=\mu$ . If $X=(\zeta,0),\quad
 Y=(\zeta ',0),\quad X,Y\in n$ then symplectic structure
  is
\begin{eqnarray*}
  \Omega (\sigma({\cal O},X)_f,\sigma({\cal
  O},Y)_f)=<f,[X,Y]>=
  f(0,B(\zeta,\zeta'))\mu B(\zeta,\zeta')
\end{eqnarray*}
Also we have for orbit ${\cal O}_\mu=N/Z$ and ${\cal O}_\mu $ is
Hamiltonian G-space.

According to this approach we
can construct by using methods of geometric quantization theory
many "symplectic wavelet constructions" with corresponding
symplectic or Poisson structure on it.
Very useful particular spline--wavelet basis with uniform
exponential control on stratified and nilpotent Lie groups
was considered in [13].

\subsection{Applications to Melnikov Functions Approach}
 We give now some point of applications
of wavelet methods from the preceding parts to Melnikov approach
in the theory of
homoclinic chaos in perturbed Hamiltonian systems for examples from [1].

In  Hamiltonian form we have:
\begin{eqnarray*}
\dot{x}=J\cdot\nabla H(x)+\varepsilon g(x,\Theta), \quad
\dot{\Theta}&=&\omega,\quad
(x,\Theta)\in R^n\times T^m,
\end{eqnarray*}
for $\varepsilon=0 $ we have:
\begin{equation}
\dot{x}=J\cdot\nabla H(x),\quad
\dot\Theta=\omega
\end{equation}
 For $\varepsilon=0$ we
have homoclinic orbit $\bar{x}_{0}(t)$ to the hyperbolic fixed
point $x_0$. For $\varepsilon\neq 0$ we have normally hyperbolic
invariant torus $T_{\varepsilon}$ and condition on transversally
intersection of stable and unstable ma\-ni\-folds
$W^s(T_{\varepsilon})$ and $W^u(T_{\varepsilon})$ in terms of
Melnikov functions $M(\Theta)$ for $\bar{x}_{0}(t)$:
 $$
M(\Theta)=\displaystyle\int\limits_{-\infty}^{\infty}\nabla
H(\bar{x}_{0}(t)) \wedge g(\bar{x}_{0}(t),\omega t+\Theta)dt
$$
This condition has the next form:
\begin{eqnarray*}
M(\Theta_0)=0, \qquad
\sum\limits_{j=1}^{2}\omega_j\frac{\partial}{\partial\Theta_j}
M(\Theta_0)\neq0
\end{eqnarray*}
According to the approach of Birkhoff-Smale-Wiggins  we
determined the region in parameter space in which we can observe the
chaotic behaviour [4].\\
 If we cannot solve equations (34)
explicitly in time, then we use the wavelet approach from paper [1]
for the computations of  homoclinic (heteroclinic) loops as
the wavelet solutions of system (34).
For computations of quasiperiodic Melnikov functions
$$
M^{m/n}(t_0)=\int^{mT}_0 DH(x_\alpha(t))\wedge g(x_\alpha(t),t+t_0)dt
$$
  we used periodization of wavelet construction from paper [1].\\
 We also used symplectic Melnikov function approach in which we have:
\begin{eqnarray*}
M_i(z)&=&\displaystyle\lim_{j\rightarrow\infty}\int\limits_{-T_j^*}
^{T_j}\{h_i,\hat{h}\}_{\Psi (t,z)}dt  \\
d_i(z,\varepsilon)&=&h_i(z^u_\varepsilon)-h_i(z^s_\varepsilon)=
\varepsilon M_i(z)+O(\varepsilon^2)
\end{eqnarray*}
where $\{,\}$ is the Poisson bracket,
$d_i(z,\varepsilon)$ is the Melnikov distance. So, we need symplectic
invariant wavelet expressions for Poisson brackets. The computations
 are produced  according to invariant calculation of Poisson brackets,
which is based on consideration in part III A and on operator representation
from part III F (see below).\\

\subsection{Floer Approach for Closed Loops}
Now we consider the generalization of  wavelet variational
approach  to the symplectic invariant calculation of
closed loops in Hamiltonian systems
[15]. As we demonstrated in [4] we have the parametrization
of our solution by some
reduced algebraical problem but in contrast to the cases from paper [1], where
the solution is parametrized by construction based on scalar
refinement equation, in symplectic case we have
parametrization of the solution
by matrix problems -- Quadratic Mirror Filters equations.
Now we consider a different approach.

Let$(M,\omega$) be a compact symplectic manifold of dimension $2n$, $\omega$ is
a closed 2-form (nondegenerate) on $M$ which induces an isomorphism $T^*M\to
TM$. Thus every smooth time-dependent Hamiltonian $H:{\bf R}\times M\to {\bf
R}$ corresponds to a time-dependent Hamiltonian vector field $X_H: {\bf
R}\times M\to TM$ defined by
\begin{equation}
\omega(X_H(t,x),\xi)=-{\rm}d_xH(t,x)\xi
\end{equation}
 for
$\xi\in T_xM$. Let $H$ (and $X_H$) is periodic in time: $H(t+T,x)=H(t,x)$ and
consider corresponding Hamiltonian differential equation on $M$:
\begin{equation}\label{eq:xdot}
\dot x(t)=X_H(t,x(t))
\end{equation}
The solutions $x(t)$ of (\ref{eq:xdot}) determine a 1-parameter
family of diffeomorphisms
$\psi_t\in {\rm Diff}(M)$ satisfying $\psi_t(x(0))=x(t)$. These diffeomorphisms are
symplectic:
 $\omega=\psi_t^*\omega$. Let $L=L_TM $ be the space of contractible loops in $M$
which are represented by smooth curves $\gamma: {\bf R}\to M$ satisfying
$\gamma(t+T)=\gamma(t)$. Then the contractible T-periodic solutions of
(\ref{eq:xdot}) can be characterized as the critical points of the functional
$S=S_T: L\to {\bf R}$:
\begin{equation}\label{eq:ST}
S_T(\gamma)=-\int_Du^*\omega+\int_0^TH(t,\gamma(t)){\rm d}t,
\end{equation}
where $D\subset {\bf C}$ be a closed unit disc and $u: D\to M$ is a smooth
function, which on boundary agrees with $\gamma$, i.e. $u({\rm exp}\{2\pi i
\Theta\})=\gamma(\Theta T)$. Because  [$\omega$], the cohomology class of
$\omega$, vanishes then
$S_T(\gamma)$ is independent of choice of $u$.
Tangent space $T_\gamma L$ is the space of vector fields $\xi\in
C^\infty(\gamma^*TM)$ along $\gamma$ satisfying $\xi(t+T)=\xi(t)$.
Then we have for the 1-form ${\rm d}f: TL\to{\bf R}$
\begin{equation}\label{eq:dST}
{\rm d} S_T(\gamma)\xi=\int_0^T(\omega(\dot\gamma,\xi)+{\rm
d}H(t,\gamma)\xi){\rm d}t
\end{equation}
and the critical points of $S$ are contractible loops in $L$ which satisfy the
Hamiltonian equation (\ref{eq:xdot}). Thus the critical points are precisely
the required T-periodic solution of (\ref{eq:xdot}).

To describe the gradient of $S$ we choose $a$ on almost complex structure on $M$
which is compatible with $\omega$. This is an endomorphism $J\in C^\infty({\rm
End}(TM))$ satisfying $J^2=-I$ such that
\begin{equation}\label{eq:fung}
g(\xi,\eta)=\omega(\xi,J(x)\eta),\qquad \xi,\eta\in T_xM
\end{equation}
defines a Riemannian metric on M. The Hamiltonian vector field is then
represented by $X_H(t,x)=J(x)\nabla H(t,x)$, where $\nabla$
denotes the gradient  w.r.t. the x-variable using the metric (\ref{eq:fung}).
Moreover the gradient of $S$ w.r.t. the induced metric on $L$ is given by
\begin{equation}
{\rm grad} S(\gamma)=J(\gamma)\dot\gamma+\nabla H(t,\gamma),\qquad
\gamma\in L
\end{equation}
Studying the critical points of $S$  is confronted with the well-known
difficulty that the variational integral is neither bounded from below nor from
above. Moreover, at every possible critical point the Hessian of $f$ has an
infinite dimensional positive and an infinite dimensional negative subspaces, so
the standard Morse theory is not applicable.
The additional problem is that the gradient vector field on the loop space $L$
\begin{equation}
\frac{\rm d}{{\rm d}s}\gamma=-{\rm grad}f(\gamma)
\end{equation}
does not define a well posed Cauchy problem.
But Floer [15] found a way to analyse the space ${\mathcal M}$ of bounded solutions
consisting of the critical points together with their connecting orbits.
He used a combination of variational approach and Gromov's elliptic technique.
A gradient flow line of $f$ is a smooth solution $u: {\bf R}\to M$ of the
partial differential equation
\begin{equation}\label{eq:duds}
\frac{\partial u}{\partial s}+J(u)\frac{\partial u}{\partial
t}+\nabla H(t,u)=0,
\end{equation}
which satisfies $u(s,t+T)=u(s,t)$. The key point is to consider (\ref{eq:duds})
not as the flow on the loop space but as an elliptic boundary value problem. It
should be noted that (\ref{eq:duds}) is a generalization of equation for
Gromov's pseudoholomorphic curves (correspond to the case $\nabla
H=0$ in (\ref{eq:duds})).
Let ${\mathcal M}_T={\mathcal M}_T(H,J)$ the space of bounded solutions of (\ref{eq:duds}), i.e. the
space of smooth functions $u: {\bf C}/ iT{\bf Z}\to M$, which are contractible,
solve equation (\ref{eq:duds}) and have finite energy flow:
\begin{equation}\label{eq:PhiT}
\Phi_T(u)=\frac{1}{2}\int_{-\infty}^\infty\int_0^T\lgroup\arrowvert\frac{\partial u}
{\partial
 s}\arrowvert^2+\arrowvert\frac{\partial u}{\partial t}-X_H(t,u)\arrowvert^2\rgroup{\rm d}t{\rm
 d}s\quad <\infty.
\end{equation}
For every $u\in M_T$ there exists a pair $x,y$ of contractible T-periodic
solutions of (\ref{eq:xdot}), such that $u$ is a connecting orbit from $y$ to
$x$:
\begin{equation}
\lim_{s\to-\infty}u(s,t)=y(t), \qquad \lim_{s\to+\infty}=x(t)
\end{equation}
Then the approach from [1], which we may apply or on the level of standard
boundary problem (42) or on the level of variational approach (43) and
representation of operators (in our case, $J$ and $\nabla$)
 according to part III F (see below) lead
us to wavelet representation of closed loops.

\subsection{Quasiclassical Evolution}

Let us consider classical and quantum dynamics in phase space
$\Omega=R^{2m}$ with coordinates $(x,\xi)$ and generated by
Hamiltonian ${\cal H}(x,\xi)\in C^\infty(\Omega;R)$.
If $\Phi^{\cal H}_t:\Omega\longrightarrow\Omega$ is (classical) flow then
time evolution of any bounded classical observable or
symbol $b(x,\xi)\in C^\infty(\Omega,R)$ is given by $b_t(x,\xi)=
b(\Phi^{\cal H}_t(x,\xi))$. Let $H=Op^W({\cal H})$ and $B=Op^W(b)$ are
the self-adjoint operators or quantum observables in $L^2(R^n)$,
representing the Weyl quantization of the symbols ${\cal H}, b$ [14]
\begin{eqnarray*}
(Bu)(x)=\frac{1}{(2\pi\hbar)^n}\int_{R^{2n}}b\left(\frac{x+y}{2},\xi\right)
\cdot
e^{i<(x-y),\xi>/\hbar}u(y)\ud y\ud\xi,
\end{eqnarray*}
where $u\in S(R^n)$ and $B_t=e^{iHt/\hbar}Be^{-iHt/\hbar}$ be the
Heisenberg observable or quantum evolution of the observable $B$
under unitary group generated by $H$. $B_t$ solves the Heisenberg equation of
motion
$$\dot{B}_t=\frac{i}{\hbar}[H,B_t].$$
Let $b_t(x,\xi;\hbar)$ is a symbol of $B_t$ then we have
 the following equation for it
\begin{equation}
\dot{b}_t=\{ {\cal H}, b_t\}_M,
\end{equation}
with initial condition $b_0(x,\xi,\hbar)$ $=b(x,\xi)$.
Here $\{f,g\}_M(x,\xi)$ is the Moyal brackets of the observables
$f,g\in C^\infty(R^{2n})$, $\{f,g\}_M(x,\xi)=f\sharp g-g\sharp f$,
where $f\sharp g$ is the symbol of the operator product and is presented
by the composition of the symbols $f,g$
\begin{eqnarray*}
(f\sharp g)(x,\xi)=&&\frac{1}{(2\pi\hbar)^{n/2}}\int_{R^{4n}}
e^{-i<r,\rho>/\hbar+i<\omega,\tau>/\hbar}
\cdot f(x+\omega,\rho+\xi)\cdot\\
&&g(x+r,\tau+\xi)\ud\rho \ud\tau \ud r\ud\omega.
\end{eqnarray*}
For our problems it is useful that $\{f,g\}_M$ admits the formal
expansion in powers of $\hbar$:
$$\{f,g\}_M(x,\xi)\sim \{f,g\}+2^{-j}
\sum_{|\alpha+\beta|=j\geq 1}(-1)^{|\beta|}\cdot
(\partial^\alpha_\xi fD^\beta_x g)\cdot(\partial^\beta_\xi
 gD^\alpha_x f),
$$
 where $\alpha=(\alpha_1,\dots,\alpha_n)$ is
a multi-index, $|\alpha|=\alpha_1+\dots+\alpha_n$,
$D_x=-i\hbar\partial_x$.
So, evolution (45) for symbol $b_t(x,\xi;\hbar)$ is
\begin{eqnarray}
\dot{b}_t=\{{\cal H},b_t\}+\frac{1}{2^j}
\sum_{|\alpha|+\beta|=j\geq 1}(-1)^{|\beta|}
\cdot\hbar^j
(\partial^\alpha_\xi{\cal H}D_x^\beta b_t)\cdot
(\partial^\beta_\xi b_t D_x^\alpha{\cal H}).
\end{eqnarray}

At $\hbar=0$ this equation transforms to classical Liouville equation
\begin{equation}
\dot{b}_t=\{{\cal H}, b_t\}.
\end{equation}
Equation (46) plays a key role in many quantum (semiclassical) problem.
We note only the problem of relation between quantum and classical evolutions
or how long the evolution of the quantum observables is determined by the
corresponding classical one [14].
Our approach to solution of systems (46), (47) is based on our technique
from [1]-[7] and very useful linear parametrization for differential operators
which we present in section III F.

\subsection{SYMPLECTIC HILBERT SCALES VIA WA\-VE\-LETS}

We can solve many important dynamical problems such that KAM
perturbations, spread of energy to higher modes, weak turbulence, growths of
solutions of Hamiltonian equations only if we consider scales of spaces instead
of one functional space. For Hamiltonian system and their perturbations
for which we need take into account underlying symplectic structure we
need to consider symplectic scales of spaces. So, if
$\dot{u}(t)=J\nabla K(u(t))$
is Hamiltonian equation we need wavelet description of symplectic or
quasicomplex structure on the level of functional spaces. It is very
important that according to [16] Hilbert basis is in the same time a
Darboux basis to corresponding symplectic structure.
We need to provide Hilbert scale $\{Z_s\}$ with symplectic structure [16], [17].
All what we need is the following.
 $J$  is a linear operator, $J : Z_{\infty}\to Z_\infty$,
$J(Z_\infty)=Z_\infty$, where $Z_\infty =\cap Z_s$.
$J$ determines an isomorphism of  scale $\{Z_s\}$ of order $d_J\geq 0$.
 The operator $J$ with domain of definition $Z_\infty$ is
antisymmetric in $Z$:
$
<J z_1,z_2>_Z=-<z_1,J z_2>_Z, z_1,z_2 \in $ $ Z_\infty
$.
Then the triple
$$\{Z,\{Z_s|s\in R\},\quad
\alpha=<\bar J dz,dz>\}
$$
 is symplectic Hilbert scale. So, we may consider
any dynamical Hamiltonian problem on functional level.
As an example, for KdV equation we have
$$ Z_s=\{u(x)\in H^s(T^1)|\int^{2\pi}_0 u(x)\ud x=0\},\
s\in R,\quad J=\partial/\partial x,$$
J is isomorphism  of the scale of  order one, $\bar J=-(J)^{-1}$ is
isomorphism of order $-1$.
According to [18] general functional spaces and scales of spaces such as
Holder--Zygmund, Triebel--Lizorkin and Sobolev can be characterized
through wavelet coefficients or wavelet transforms. As a rule, the faster
the wavelet coefficients decay, the more the analyzed function is
regular [18]. Most important for us example is the scale of Sobolev spaces.
Let $H_k(R^n)$ is the Hilbert space of all distributions with finite norm
$$
\Vert s\Vert^2_{H_k(R^n)}=\int \ud\xi(1+\vert\xi\vert^2)^{k/2}\vert
\hat s(\xi)\vert^2.
$$
Let us consider wavelet transform
$$
W_g f(b,a)=\int_{R^n}\ud x\frac{1}{a^n}\bar g\left(\frac{x-b}{a}\right) f(x),
$$
$ b\in R^n, \quad a>0$,
w.r.t. analyzing wavelet $g$, which is strictly admissible, i.e.
$$
C_{g,g}=\int_0^\infty\frac{\ud a}{a}\vert\bar{\hat g(ak)}\vert^2<\infty.
$$
Then there is a $c\geq 1$ such that
\begin{eqnarray*}
c^{-1}\Vert s \Vert^2_{H_k(R^n)}\leq\int_{H^n}
\frac{\ud b\ud a}{a}(1+a^{-2\gamma})\vert W_gs(b,a)\vert^2
\leq
c\|s\|^2_{H_k(R^n)}.
\end{eqnarray*}
This shows that localization of the wavelet coefficients at small
scale is linked to local regularity.

So, we need representation for differential operator ($J$ in our case) in
wavelet basis. We consider it in the next section.

\subsection{FAST WAVELET TRANSFORM FOR DIF\-FE\-RENTIAL OPERATORS}

Let us consider multiresolution representation
$\dots\subset V_2\subset V_1\subset V_0\subset V_{-1}
\subset V_{-2}\dots$ (see our other paper from this proceedings for
details of wavelet machinery). Let T be an operator $T:L^2(R)
\rightarrow L^2(R)$, with the kernel $K(x,y)$ and
$P_j: L^2(R)\rightarrow V_j$ $(j\in Z)$ is projection
operators on the subspace $V_j$ corresponding to j level of resolution:
$$(P_jf)(x)=\sum_k<f,\varphi_{j,k}>\varphi_{j,k}(x).$$ Let
$Q_j=P_{j-1}-P_j$ is the projection operator on the subspace $W_j$ then
we have the following "microscopic or telescopic"
representation of operator T which takes into account contributions from
each level of resolution from different scales starting with
coarsest and ending to finest scales:
$$
T=\sum_{j\in Z}(Q_jTQ_j+Q_jTP_j+P_jTQ_j).
$$
We remember that this is a result of presence of affine group inside this
construction.
The non-standard form of operator representation [19] is a representation of
an operator T as  a chain of triples
$T=\{A_j,B_j,\Gamma_j\}_{j\in Z}$, acting on the subspaces $V_j$ and
$W_j$:
$$
 A_j: W_j\rightarrow W_j, B_j: V_j\rightarrow W_j,
\Gamma_j: W_j\rightarrow V_j,
$$
where operators $\{A_j,B_j,\Gamma_j\}_{j\in Z}$ are defined
as $$A_j=Q_jTQ_j, \quad B_j=Q_jTP_j, \quad\Gamma_j=P_jTQ_j.$$
The operator $T$ admits a recursive definition via
$$T_j=
\left(\begin{array}{cc}
A_{j+1} & B_{j+1}\\
\Gamma_{j+1} & T_{j+1}
\end{array}\right),$$
where $T_j=P_jTP_j$ and $T_j$ works on $ V_j: V_j\rightarrow V_j$.
It should be noted that operator $A_j$ describes interaction on the
scale $j$ independently from other scales, operators $B_j,\Gamma_j$
describe interaction between the scale j and all coarser scales,
the operator $T_j$ is an "averaged" version of $T_{j-1}$.

The operators $A_j,B_j,\Gamma_j,T_j$ are represented by matrices
$\alpha^j, \beta^j, \gamma^j, s^j$
\begin{eqnarray}
\alpha^j_{k,k'}&=&\int\int K(x,y)\psi_{j,k}(x)\psi_{j,k'}(y)\ud x\ud y\nonumber\\
\beta^j_{k,k'}&=&\int\int K(x,y)\psi_{j,k}(x)\varphi_{j,k'}(y)\ud x\ud y\\
\gamma^j_{k,k'}&=&\int\int K(x,y)\varphi_{j,k}(x)\psi_{j,k'}(y)\ud x\ud y\nonumber\\
s^j_{k,k'}&=&\int\int K(x,y)\varphi_{j,k}(x)\varphi_{j,k'}(y)\ud x\ud y\nonumber
\end{eqnarray}
We may compute the non-standard representations of operator $\ud/\ud x$ in the
wavelet bases by solving a small system of linear algebraical
equations. So, we have for objects (48)
\begin{eqnarray*}
\alpha^j_{i,\ell}&=&2^{-j}\int\psi(2^{-j}x-i)\psi'(2^{-j}-\ell)2^{-j}\ud x
=2^{-j}\alpha_{i-\ell}\\
\beta^j_{i,\ell}&=&2^{-j}\int\psi(2^{-j}x-i)\varphi'(2^{-j}x-\ell)2^{-j}\ud x
=2^{-j}\beta_{i-\ell}\\
\gamma^j_{i,\ell}&=&2^{-j}\int\varphi(2^{-j}x-i)\psi'(2^{-j}x-\ell)2^{-j}\ud x
=2^{-j}\gamma_{i-\ell},
\end{eqnarray*}
where
\begin{eqnarray*}
\alpha_\ell&=&\int\psi(x-\ell)\frac{\ud}{\ud x}\psi(x)\ud x\\
\beta_\ell&=&\int\psi(x-\ell)\frac{\ud}{\ud x}\varphi(x)\ud x\\
\gamma_\ell&=&\int\varphi(x-\ell)\frac{\ud}{\ud x}\psi(x)\ud x
\end{eqnarray*}
then by using refinement equations
\begin{eqnarray*}
\varphi(x)&=&\sqrt{2}
\sum^{L-1}_{k=0}h_k\varphi(2x-k),\\
 \psi(x)&=&\sqrt{2}\sum_{k=0}^{L-1}g_k\varphi(2x-k),
\end{eqnarray*}
$g_k=(-1)^kh_{L-k-1},\quad k=0,\dots,L-1$ we have in terms of filters
$(h_k,g_k)$:
\begin{eqnarray*}
\alpha_j&=&2\sum^{L-1}_{k=0}\sum^{L-1}_{k'=0}g_kg_{k'}r_{2i+k-k'},\\
\beta_j&=&2\sum^{L-1}_{k=0}\sum^{L-1}_{k'=0}g_kh_{k'}r_{2i+k-k'},\\
\gamma_i&=&2\sum^{L-1}_{k=0}\sum^{L-1}_{k'=0}h_kg_{k'}r_{2i+k-k'},
\end{eqnarray*}
where $$r_\ell=\int\varphi(x-\ell)\frac{\ud}{\ud x}\varphi(x)\ud x, \ell\in Z.$$
Therefore, the representation of $d/dx$ is completely determined by the
coefficients $r_\ell$ or by representation of $d/dx$ only on
the subspace $V_0$. The coefficients $r_\ell, \ell\in Z$ satisfy the
following system of linear algebraical equations
$$
r_\ell=2\left[ r_{2l}+\frac{1}{2} \sum^{L/2}_{k=1}a_{2k-1}
(r_{2\ell-2k+1}+r_{2\ell+2k-1}) \right]
$$
and $\sum_\ell\ell r_\ell=-1$, where $a_{2k-1}=$
$2\sum_{i=0}^{L-2k}h_i h_{i+2k-1}$, $k=1,\dots,L/2$
are the autocorrelation coefficients of the filter $H$.
If a number of vanishing moments $M\geq 2$ then this linear system of equations
has a unique solution with finite number of non-zero $r_\ell$,
$r_\ell\ne 0$ for $-L+2\leq\ell\leq L-2, r_\ell=-r_{-\ell}$.
For the representation of operator $d^n/dx^n$ we have the similar reduced
linear system of equations.
Then finally we have for action of operator $T_j(T_j:V_j\rightarrow V_j)$
on sufficiently smooth function $f$:
$$
(T_j f)(x)=\sum_{k\in Z}\left(2^{-j}\sum_{\ell}r_\ell f_{j,k-\ell}\right)
\varphi_{j,k}(x),
$$
where $\varphi_{j,k}(x)=2^{-j/2}\varphi(2^{-j}x-k)$ is wavelet basis and
$$
f_{j,k-1}=2^{-j/2}\int f(x)\varphi(2^{-j}x-k+\ell)\ud x
$$
are wavelet coefficients. So, we have simple linear para\-met\-rization of
matrix representation of our differential operator in wavelet basis
and of the action of
this operator on arbitrary vector in our functional space.
Then we may use such representation in all preceding sections.

\section{Maps and Wavelet Structures}
\subsection{Veselov-Marsden Discretization}
Discrete variational principles lead to evolution dynamics analogous to the
Euler-Lagrange equations [9]. Let $Q$ be a configuration space, then a discrete
Lagrangian is a map $L: Q\times Q\to{\bf R}$. usually $L$ is obtained by
approximating the  given Lagrangian. For $N\in N_+$ the action sum is the map
$S: Q^{N+1}\to{\bf R}$ defined by
\begin{equation}
S=\sum_{k=0}^{N-1}L(q_{k+1}, q_k),
\end{equation}
 where
$q_k\in Q$, $k\ge 0$. The action sum is the discrete analog of the action
integral in continuous case. Extremizing $S$ over $q_1,...,q_{N-1}$ with fixing
$q_0, q_N$ we have the discrete Euler-Lagrange equations (DEL):
\begin{equation}
D_2L(q_{k+1}, q_k)+D_1(q_k, q_{q-1})=0,
\end{equation}
for $k=1,...,N-1$.

Let
\begin{equation}
\Phi: Q\times Q\to Q\times Q
\end{equation}
 and
\begin{equation}
\Phi(q_k, q_{k-1})=(q_{k+1}, q_k)
\end{equation}
 is a discrete function (map), then we have for DEL:
\begin{equation}
D_2L\circ\Phi+D_1L=0
\end{equation}
or in coordinates $q^i$ on $Q$ we have DEL
\begin{equation}
\frac{\partial L}{\partial q^i_k}\circ\Phi(q_{k+1},q_k)+
\frac{\partial L}{\partial q_{k+1}^i}(q_{k+1},q_k)=0.
\end{equation}
It is very important that the map $\Phi$ exactly preserves the symplectic form
$\omega$:
\begin{equation}
\omega=\frac{\partial^2 L}{\partial q_k^i\partial q_{k+1}^j}(q_{k+1},q_k){\rm
d}q_k^i\wedge{\rm d}q^j_{k+1}
\end{equation}

\subsection{Generalized Wavelet Approach}
Our approach to solutions of equations (54) is based on applications of general and very efficient methods
developed by A.~Harten [20], who produced
a "General Framework" for multiresolution representation of discrete data. It is
based on consideration of basic operators, decimation and prediction, which
connect adjacent resolution levels. These operators are constructed from two
basic blocks: the discretization and reconstruction operators. The former
obtains discrete information from a given continuous functions (flows), and
the latter produces an approximation to those functions, from discrete values,
in the same function space to which the original function belongs. A "new
scale" is defined as the information on a given resolution level which cannot
be predicted from discrete information at lower levels. If the discretization
and reconstruction are local operators, the concept of "new scale" is also
local. The scale coefficients are directly related to the prediction errors,
and thus to the reconstruction procedure. If scale coefficients are small at a
certain location on a given scale, it means that the reconstruction procedure on
that scale gives a proper approximation of the original function at that
particular location.
This approach may be considered as some generalization of standard wavelet
analysis approach. It allows to consider multiresolution decomposition when
usual approach is impossible ($\delta$-functions case).

Let $F$ be a linear space of mappings
\begin{equation}\label{eq:Fin}
F\subset \{f|f: X\to Y\},
\end{equation}
where $X,Y$
are linear spaces. Let also $D_k$ be a linear operator
\begin{equation}
D_k: f\to\{v^k\},\quad
 v^k=D_kf, \quad  v^k=\{v_i^k\}, \quad v_i^k\in Y.
\end{equation}
This sequence corresponds to $k$ level discretization of $X$.
Let
\begin{equation}\label{eq:Dk}
D_k(F)=V^k={\rm span}\{\eta^k_i\}
\end{equation}
and the coordinates of $v^k\in V^k$ in this basis are
$\hat{v}^k=\{\hat{v}^k_i\}$, $\hat{v}^k\in S^k$:
\begin{equation}
v^k=\sum_i\hat{v}^k_i\eta^k_i,
\end{equation}\label{eq:vk}
$D_k$ is a discretization operator.
Main goal is to design a multiresolution scheme (MR) [20] that applies to all
sequences $s\in S^L$, but corresponds for those sequences $\hat{v}^L\in S^L$,
which are obtained by the discretization (\ref{eq:Fin}).

Since $D_k$ maps $F$ onto $V^k$ then for any $v^k\subset V^k$ there is  at least
one $f$ in $F$ such that $D_kf=v^k$. Such correspondence from $f\in F$ to
$v^k\in V^k$ is reconstruction and the corresponding operator is the
reconstruction operator $R_k$:
\begin{equation}
R_k: V_k\to F, \qquad D_kR_k=I_k,
\end{equation}
where $I_k$ is the identity operator in $V^k$ ($R^k$ is right inverse of $D^k$
in $V^k$).

Given a sequence of discretization $\{D_k\}$ and sequence of the corresponding
reconstruction operators $\{R_k\}$, we define the operators $D_k^{k-1}$ and
$P^k_{k-1}$
\begin{eqnarray}
D_k^{k-1}&=&D_{k-1}R_k: V_k\to V_{k-1}\\
P^k_{k-1}&=&D_kR_{k-1}: V_{k-1}\to V_k\nonumber
\end{eqnarray}
If the set ${D_k}$ in nested [20], then
\begin{equation}
D_k^{k-1}P^k_{k-1}=I_{k-1}
\end{equation}
and we have for any $f\in F$ and any $p\in F$ for which the reconstruction
$R_{k-1}$ is exact:
\begin{eqnarray}\label{eq:DP}
D_k^{k-1}(D_kf)&=&D_{k-1}f\\
P^k_{k-1}(D_{k-1}p)&=&D_kp\nonumber
\end{eqnarray}
Let us consider any $v^L\in V^L$, Then there is $f\in F$ such that
\begin{equation}
v^L=D_Lf,
\end{equation}
and it follows from (\ref{eq:DP}) that the process of successive decimation [20]
\begin{equation}
v^{k-1}=D_k^{k-1}v^k, \qquad k=L,...,1
\end{equation}
yields for all $k$
\begin{equation}
v^k=D_kf
\end{equation}
Thus the problem of prediction, which is associated with the corresponding MR
scheme, can be stated  as a problem of approximation: knowing $D_{k-1}f$,
$f\in F$, find a "good approximation" for $D_k f$.
It is very important that each space $V^L$ has a multiresolution basis
\begin{equation}
\bar{B}_M=\{\bar{\phi}_i^{0,L}\}_i, \{\{\bar{\psi}^{k,L}_j\}_j\}^L_{k=1}
\end{equation}
and that any $v^L\in V^L$ can be written as
\begin{equation}\label{eq:vL}
v^L=\sum_i\hat{v}_i^0\bar{\phi}_i^{0,L}+\sum^L_{k=1}\sum_j
d_j^k\bar{\psi}_j^{k,L},
\end{equation}
where $\{d_j^k\}$ are the $k$ scale coefficients of the associated MR,
$\{\hat{v}_i^0\}$ is defined by (59) with $k=0$.
If $\{D_k\}$ is a nested sequence of discretization [20] and $\{R_k\}$ is any
corresponding sequence of linear reconstruction operators, then we have from
(\ref{eq:vL}) for $v^L=D_Lf$ applying $R_L$:
\begin{equation}\label{eq:RlDl}
R_LD_Lf=\sum_i \hat{f}^0_i\phi^{0,L}_i+\sum_{k=1}^L\sum_jd_j^k\psi_j^{k,L},
\end{equation}
where
\begin{equation}
\phi_i^{0,L}=R_L\bar{\phi}_i^{0,L}\in F,\quad
\psi_j^{k,L}=R_L\bar{\psi}_j^{k,L}\in F,\quad
D_0 f=\sum\hat{f}^0_i\eta^0_i.
\end{equation}
When $L\to\infty$ we have sufficient conditions which ensure that the limiting
process $L\to\infty$ in (\ref{eq:RlDl}, 70) yields a multiresolution basis for $F$.
Then, according to (67), (68) we have very useful representation for solutions
of equations (54) or for different maps construction in the form which are a
counterparts for discrete (difference) cases of constructions from paper [1].

\section*{acknowledgements}
We would like to thank Professors M.~Cornacchia, C.~Pellegrini, L.~Palumbo,
Mrs. M.~Laraneta, J.~Kono, G.~Nanula for hospitality, help, support before and
during Arcidosso meeting and all participants for interesting discussions.
We are very grateful to Professor M.~Cornacchia,
Mrs. J.~Kono and M.~Laraneta, because without their permanent encouragement this paper
would not have been written.

\end{document}